\documentclass[%
 reprint,
 amsmath,amssymb,
 aps,
]{revtex4-1}

\usepackage{color}

\usepackage{graphicx}% Include figure files
\usepackage{dcolumn}% Align table columns on decimal point
\usepackage{bm}% bold math
\begin{document}

\preprint{APS/123-QED}

%%%%%%%%%%%%%%%%%%%%%%%%%%%%%%%%%%%%%%%%%%%%%%%

\title{Influence of a dispersion of magnetic and non-magnetic nanoparticles on the magnetic Fredericksz transition of 5CB}% Force line breaks with \\

\author{Ahmed Mouhli}
\affiliation{Universit\'e de Tunis El Manar, Facult\'e des Sciences de Tunis, LR99ES16 Laboratoire de Physique de la Mati\`ere Molle et de la Mod\'elisation Electromagn\'etique (LP3ME), 2092, Tunis, Tunisie.}
\author{J\'er\^ome Fresnais}
\affiliation{Sorbonne Universit\'es, UPMC Univ Paris 06, UMR 8234, PHENIX, F75005 Paris, France. }
\author{Habib Ayeb}
\affiliation{Universit\'e de Tunis El Manar, Facult\'e des Sciences de Tunis, LR99ES16 Laboratoire de Physique de la Mati\`ere Molle et de la Mod\'elisation Electromagn\'etique (LP3ME), 2092, Tunis, Tunisie.}
\author{Ian R. Nemitz}
\affiliation{Dept of Physics, Case Western Reserve University, Cleveland, Ohio USA 44106. }
\author{Joel S. Pendery}
\affiliation{Dept of Physics, Case Western Reserve University, Cleveland, Ohio USA 44106. }
\author{Olivier Sandre}
\affiliation{Univ. Bordeaux, Bordeaux INP, CNRS UMR5629, Laboratoire de Chimie des Polym\`eres Organiques, F33607 Pessac, France}
\author{Tahar Othman}
\affiliation{Universit\'e de Tunis El Manar, Facult\'e des Sciences de Tunis, LR99ES16 Laboratoire de Physique de la Mati\`ere Molle et de la Mod\'elisation Electromagn\'etique (LP3ME), 2092, Tunis, Tunisie.}
\author{Charles Rosenblatt}
\affiliation{Dept of Physics, Case Western Reserve University, Cleveland, Ohio USA 44106. }
\author{Vincent Dupuis}
\affiliation{Sorbonne Universit\'es, UPMC Univ Paris 06, UMR 8234, PHENIX, F75005 Paris, France. }
\author{Emmanuelle Lacaze}
\email{emmanuelle.lacaze@insp.jussieu.fr}
\affiliation{Sorbonne Universit\'es, UPMC Univ Paris 06, CNRS UMR 7588, Institut des Nano-Sciences de Paris (INSP), F75005 Paris, France}

\date{\today}% It is always \today, today,

\begin{abstract}
Long time ago, Brochard and de Gennes predicted the possibility of significantly decreasing the critical magnetic field of the  Fredericksz transition (the magnetic  Fredericksz threshold) in a mixture of nematic liquid crystals and ferromagnetic particles, the so-called ferronematics. This phenomenon has rarely been measured, usually due to soft homeotropic anchoring induced at the nanoparticle surface. Here we present an optical study of the magnetic Fredericksz transition \textcolor{blue}{combined with} a light scattering study of the classical nematic liquid crystal, 5CB, doped with 6 nm diameter magnetic and non-magnetic nanoparticles. Surprisingly, for both nanoparticles, we observe at room temperature a net decrease of the threshold field of the Fredericksz transition at low nanoparticle concentrations, which appears associated with a coating of the nanoparticles by a brush of polydimethylsiloxane copolymer chains inducing planar anchoring of the director on the nanoparticle surface. Moreover the magnetic Fredericksz threshold exhibits non-monotonic behaviour as a function of the nanoparticle concentration for both types of nanoparticles, first decreasing down to a value from 23\% to 31\% below that of pure 5CB, then increasing with a further increase of nanoparticle concentration. This is interpreted as an aggregation starting at around 0.02 weight fraction that consumes more isolated nanoparticles than those introduced when the concentration is increased above $c = 0.05$ weight fraction (volume fraction $3.5 \times 10^{-2}$). This shows the larger effect of isolated nanoparticles on the threshold with respect to aggregates. From dynamic light scattering measurements we deduced that, if the decrease of the magnetic threshold when the nanoparticle concentration increases is similar for both kinds of nanoparticles, the origin of this decrease is different for magnetic and non-magnetic nanoparticles. For non-magnetic nanoparticles, the behavior may be associated with a decrease of the elastic constant due to weak planar anchoring. For magnetic nanoparticles there are non-negligible local magnetic interactions between liquid crystal molecules and magnetic nanoparticles, leading to an increase of the average order parameter. This magnetic interaction thus favors an easier liquid crystal director rotation in the presence of external magnetic field, able to reorient the magnetic moments of the nanoparticles along with the molecules.
\end{abstract}

\pacs{Valid PACS appear here}% PACS, the Physics and Astronomy
\maketitle

%%%%%%%%%%%%%%%%%%%%%%%%%%%%%%%%%%%%%%%%%%%%%%%
%
% INTRODUCTION
%
\section{Introduction}

The dielectric and diamagnetic properties of liquid crystals allow the control of their optical properties using electric or magnetic fields. However, due to their weak diamagnetism, most liquid crystal devices are mainly driven by electric fields. Mixtures of liquid crystals and ferronematic colloidal particles, the so-called ferronematics, have been the subject of a number of experimental and theoretical studies, \textcolor{blue}{related to} modifications of liquid crystal magnetic \cite{ Kopcansky1997_JP, Cruz2005, Kopcansky2010, Podoliak2012, Prodanov2016} and electrical properties \cite{Potocova1999, Kopcansky2004, Kopcansky2005, Kopcansky2006, Kopcansky2008} in the presence of a magnetic field and the recent evidence of possible formation of ferromagnetic behavior in the presence of magnetic \textcolor{blue}{nano}-platelets \cite{Mertelj2013}. One main idea of ferronematics is to enhance the weak diamagnetic response of liquid crystals, which is associated with a large magnetic Fredericksz threshold field, of the order of 1-10 kG. The first work of Brochard and de Gennes theoretically  showed that, despite their low and diamagnetic susceptibility, the magnetic field threshold of Fredericksz transition could be significantly reduced by manipulation of the nematic director using ferromagnetic nanorods, which reorient under the application of an external magnetic field \cite{Brochard1970_JP}. A number of measurements have combined dielectric capacitance measurement and application of a magnetic fields either parallel or perpendicular to the cells \cite{Potocova1999, Kopcansky2004, Kopcansky2005, Kopcansky2006, Kopcansky2008, Makarov2013}, usually allowing for the extraction of the energy of anchoring of liquid crystal molecules on the magnetic nanoparticles surface \cite{Burylov1995}. The number of direct measurements of the magnetic Fredericksz threshold field in the presence of magnetic nanoparticles is scarce. Yet it was shown that the threshold field decreases (respectively increases) with concentration of dispersed magnetic nanoparticles when the sign of the anisotropy of magnetic susceptibility the raw nematic is negative (respectively positive). The maximum decrease with respect to the pure liquid crystal has been found to be of the order of 50\% \cite{Kopcansky1997_JP, Kopcansky2010, Kopcansky2013}, and recent measurements at room temperature have revealed a decrease of Fredericksz threshold of 35\% for magnetic nanospheres covered by dendrimers and dispersed in E7 \cite{Prodanov2016}. The origin of the experimental results has been explained by a number of theoretical works \cite{Burylov1990, Burylov1993_JMMM, Burylov1995, Burylov1995B, Burylov2013}, enlarging on the Brochard and de Gennes theory to include the assumption of non-rigid anchoring. Mainly homeotropic anchoring and negative magnetic anisotropies haves been considered, leading to an increase of the magnetic threshold field instead of a decrease \cite{Burylov1990, Burylov1993_JMMM, Burylov1995B, Makarov2010}, consistent with a number of experimental results \cite{Chen1983, Kopcansky1997_JP, Kopcansky2010}.  However, planar anchoring has been studied also in detail recently, leading to the prediction of a decrease of the magnetic threshold \cite{Zakhlevnykh2014, Zakhlevnykh2016}. Moreover only nanorods or platelets have been theoretically considered \cite{Gorkunov2011, Burylov2013, Raikher2013_SM}. A departure with respect to the theory was observed when the condition of small concentration of nanoparticles could no more be considered as valid, the departure point being estimated at volume fraction near $2\times 10^{-4}$ \cite{Podoliak2012}.

 Here we show magnetic Freedericksz transition feature for the classical nematic liquid crystal, 5CB (4′-pentyl-4-biphenylcarbonitrile) in the presence of magnetic (iron oxide) and non-magnetic (cerium oxide) nanospheres with planar anchoring for a range of concentrations, including concentrations well above the previously estimated non-aggregation limit (volume fraction between $2 \times 10^{-3}$ and $4.5 \times 10^{-2}$, the citrate ligand and polydimethysiloxane (PDMS) shell being included in the calculation of volume fraction). We \textcolor{blue}{establish the curve of the magnetic Freedericksz threshold evolution with the nanoparticle concentration, up to now only rarely presented in the litterature. It leads to the observation of} two concentration ranges, one at low concentrations where the influence of single (i.e. individually dispersed) nanoparticles dominates, the second at large nanoparticle concentrations where the influence of aggregates dominates, finally allowing for an unambiguous description of the behavior induced by single nanoparticles. We demonstrate a surprisingly large effect for both magnetic and non-magnetic nanoparticles, whereas \textcolor{blue}{this latter} phenomenon had been shown previously with nonmagnétic dopants (carbon nanotubes) but only for low magnetic field values \cite{Tomasovicova2013}. We combine measurements of the magnetic Fredericksz threshold together with dynamic light scattering (DLS) measurements and evidence the microscopic behavior associated with the observed magnetic threshold decrease of 23\% at room temperature for non-magnetic nanoparticles and of 31\% for magnetic nanoparticles, in relation with the planar anchoring of 5CB induced by the PDMS shell coating on the nanoparticles. However, a different influence of both nanoparticles is found concerning the liquid crystal director profile in the nanoparticle vicinity, allowing us to decipher an alternative mechanism as the origin of the threshold field reduction. The magnetic interactions  between nanoparticles and liquid crystal molecules are found to be far from negligible, which could constitute a crucial feature on which to build future theories, including magnetic nanospheres rarely considered in theoretical works \cite{Zadorozhnii2006}. These spherical nanoparticles grafted by a polymer brush are arguably the most interesting dopants for applications, as they are far easier to synthesize than magnetic nanorods.

%%%%%%%%%%%%%%%%%%%%%%%%%%%%%%%%%%%%%%%%%%%%%%%

\section{Experimental}

\subsection{Samples, cell preparation}
\begin{figure}[h]
\centering
  \includegraphics[width=8cm]{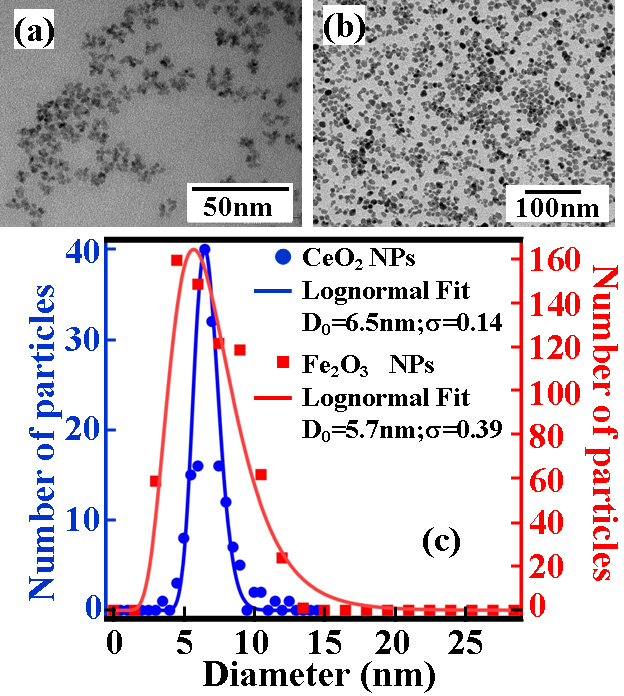}
  \caption{TEM images of cerium oxide (a) and iron oxide (c) nanoparticles; size distributions from TEM analysis (b)}
  \label{fgr:TEM}
  \end{figure}
  
\begin{figure}[h]
\centering
  \includegraphics[width=8cm]{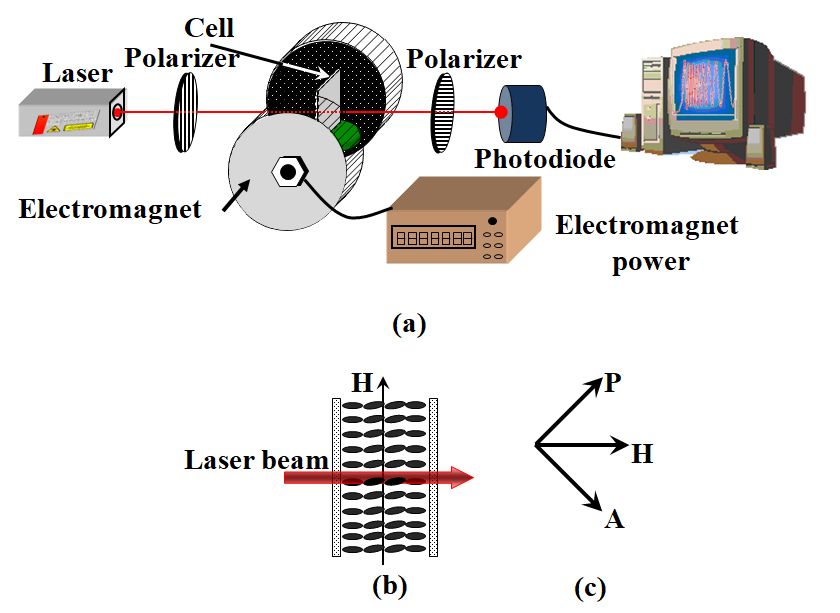}
  \caption{Experimental setup: a linearly polarized red laser beam passes trough the sample submitted to an in-plane magnetic field and is analysed by a polarizer and a photodiode; sketch of a sample cell submitted to a magnetic field H showing the competition between homeotropic anchoring on the cell walls and the magnetic alignment force.}
  \label{fgr:setup1}
  \end{figure}
	5CB (4-pentyl-4-biphenylcarbonitrile) liquid crystal and Trichloro(1H,1H,2H,2H-perfluorooctyl)silane were purchased from Sigma-Aldrich and used as received. Poly(amninopropylmethylsiloxane-$b$-dimethylsiloxane),thereafter called PAPMS-b-PDMS block copolymer, was purchased from Gelest Inc. USA and used as received. Superparamagnetic nanoparticles ($\gamma Fe_{2}O_{3}$) were synthesized using the Massart method \cite{massart1981preparation, massart1995preparation}. Post-synthesis size sorting was applied to obtain lower size dispersity \cite{Bee1995}.  Non-magnetic cerium oxide nanoparticles ($CeO_{2}$) were synthesized by thermo-hydrolysis of cerium nitrate salt provided by Solvay (Centre de Recherche d'Aubervilliers, Aubervilliers, France) under hydrothermal conditions at neutral or acidic pH. Transmission electron microscopy (TEM) pictures performed on dilute dispersions  are shown on Figure \ref{fgr:TEM} for both particles types. Their analysis allows the determination of the nanoparticle mean diameter and standard deviation. Alternatively, the size distributions were fitted by lognormal functions. Cerium oxide nanoparticles have a median diameter $d_{0} = 6.5 nm$ and polydispersity $\sigma = 0.14$. Iron oxide nanoparticles have a median diameter $d_{0} = 5.7 nm$ and polydispersity $\sigma = 0.39$ (figure \ref{fgr:TEM}). The mean diameters can be deduced by the formula $d = d_{0} exp(\sigma^{2}/2)$, giving average diameters respectively of $6.6 nm$ for cerium oxide and $6.2 nm$ for iron oxide. The particles were functionnalized first with a citrate coating, then coated by the PAPMS-b-PDMS block copolymer and dispersed in dichloromethane, which consists in a slightly modified pathway compared to previously published work \cite{sotebier2012}. Both nanoparticles are made of a metal oxide exhibiting hydroxyl groups, thus with the same propensity to be coated first by the citrate layer, then with the PAPMS-b-PDMS which forms a PDMS brush around the citrate self-assembled monolayer. The PDMS shell is expected to be of width around 2nm, 5CB being a bad solvent for the PDMS block. This leads to a difference in size not larger than 4\% between magnetic and non-magnetic nanoparticles in presence of the PDMS shell. The PDMS shell is also expected to smooth the small remaining difference in size, size polydispersity and faceting between the two nanoparticles. If the magnetic nature of the particles is ignored, the two kinds of nanoparticles behave regarding anchoring of 5CB molecules as if they were plain PDMS particles of very similar size, shape and polydispersity. 

To fabricate observation cells and achieve homeotropic anchoring of 5CB, a glass sheet purchased from SOLEMS-France was cut in pieces (20x15x1 mm) that were cleaned with acetone in an ultrasonic cleaner for 30 minutes and rinsed with distilled water. The glass slides were treated for 2 minutes in a O$_{2}$ plasma reactor (Diener Femto) at 0.5 mPa  to activate the surfaces with Si-OH moieties. Slides were placed rapidly into a desiccator at 100$^{\circ}$C with trichloro(1H,1H,2H,2H-perfluorooctyl)silane for 60 minutes in order to induce chemical bonding between glass slides and the silane deposited in the vapor phase, with the silane inducing homeotropic anchoring of 5CB because of its hydrophobic and oleophobic character.

A solution of 5CB was prepared at 1\% by weight in dichloromethane. The necessary amount of the iron oxide nanoparticles dispersed in dichloromethane was added to reach the final concentration. Dichloromethane from the mixture was evaporated at 60$^{\circ}$C on a hot plate then cooled down rapidly on ice. The mixture was placed on a silanized hydrophobic glass slide at room temperature. Then, Mylar spacers were used to ensure a 50$\mu$m gap and the second hydrophobic glass slides was placed on top to form the cell. The sample was placed in the cell holder for measurements with a final concentration of cerium oxide or iron oxide nanoparticles ranging from 0.01 to 0.15 weight fraction.
\subsection{Birefringence under magnetic field: Set-up}
\begin{figure}[h]
\centering
  \includegraphics[width=8cm]{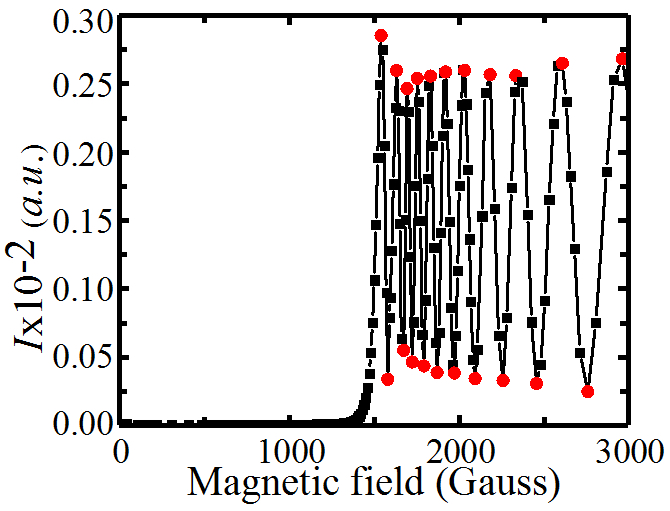}
  \caption{Typical signal collected by the photodiode. The red dots correspond to the extrema where the phase lag equals $m\pi/2$, where $m$ is an integer.}
  \label{fgr:rawdata}
\end{figure}
The experimental setup used for investigation of the Fredericksz transition in pure and nanoparticle doped liquid crystals is shown on Figure \ref{fgr:setup1}a. It consists of a computer-controlled electromagnet with a gap of 18 mm that allows applying a homogeneous magnetic field up to 1T. The magnetic field was applied in the plane of the sample cell (Figure \ref{fgr:setup1}b) to induce the Fredericksz reorientation transition, resulting in the appearance of a non-zero birefringence in the sample. To measure this birefringence, we used a He-Ne laser beam linearly polarized at 45$^\circ$ from the magnetic field direction, so that each component sees a different optical index respectively $n_{||}$ and $n_{\perp}$ (Figure \ref{fgr:setup1}c, $\Delta n = n_{||} - n_{\perp}$ is the birefringence of the sample). If the integrated birefringence over the light path is not an integer multiple of $2\pi$, the light emerging from the sample is elliptically polarized and is analyzed by a second polarizer crossed with respect to the first. The light is detected by a photodiode whose voltage is output into a computer. A typical experiment consists in the recording of the birefringence signal while slowly ramping the magnetic field.
\subsection{Birefringence under magnetic field: Data analysis}
The typical signal collected on the photodiode is shown in Figure \ref{fgr:rawdata}. It consists of a succession of peaks given by the expression for the transfer function, where $e$ is the cell thickness,

\begin{equation}
I(H) = I_0 \sin^2\left(\frac{2\pi e\Delta n(H)}{2\lambda}\right) 
\end{equation}

\begin{figure}[h]
\centering
  \includegraphics[width=8cm]{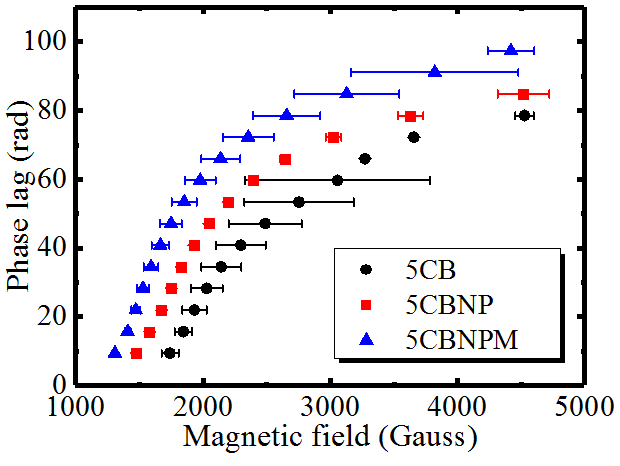}
  \caption{Integrated birefringence vs field for the pure 5CB and 0.05 wt fraction doped 5CB with non-magnetic nanoparticles (5CBNP) and magnetic nanoparticles (5CBNPM).}
  \label{fgr:biref}
\end{figure}

\begin{figure}[h]
\centering
  \includegraphics[width=8cm]{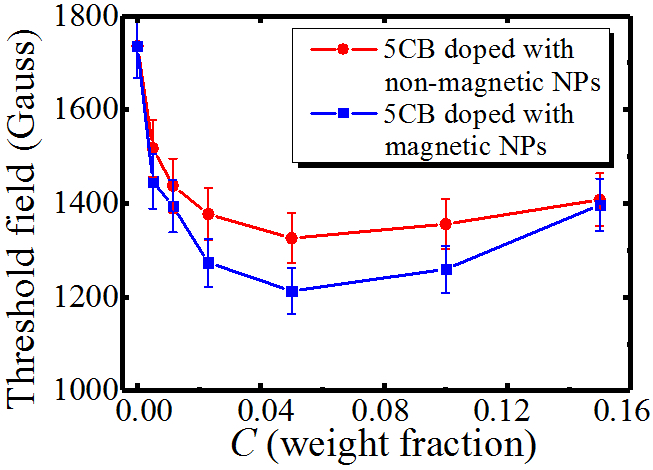}
  \caption{Evolution of the magnetic threshold field as a function of the nanoparticle weight concentration for non-magnetic and magnetic nanoparticles.}
  \label{fgr:field_concentration}
\end{figure}

Each time the argument of the sine is equal to $m \pi$, with $m$ an integer, the signal passes through an extremum. Thus it is possible to reconstruct the field dependence of the integrated birefringence step by step by locating the different field values at which the extrema appear. The corresponding integrated birefringence is presented as a function
of magnetic field in Figure. \ref{fgr:biref}, for pure 5CB (black circle), for 5CB with non-magnetic nanoparticles (red square, concentration 0.05 weight fraction) and for 5CB with magnetic nanoparticles (blue triangle, concentration 0.05 weight fraction). The signal is at first nearly flat and equal to zero, then birefringence starts to rise continously with the field intensity, before reaching a plateau value. The threshold of the magnetic field, $H_{c}$, above which the birefringence changes (magnetic Fredericksz transition threshold), is determined through an extrapolation of the rising curve and the determination of its intersection with the X axis. The evolution of $H_{c}$ is presented in Figure. \ref{fgr:field_concentration} as a function of the nanoparticle weight concentration, for non-magnetic nanoparticles (cerium oxide nanoparticles-red circle) and for
magnetic nanoparticles (iron oxide nanoparticles-blue square). We observe that not only magnetic nanoparticles, but also non-magnetic nanoparticles, influence
 $H_{c}$: First a decrease of the magnetic threshold is observed with increasing nanoparticle concentration until a minimum threshold magnetic field is reached for a concentration around
0.05 weight fraction. $H_{c}$ then increases with increasing concentration.

\subsection{Dynamic light scattering measurement (DLS)}
\begin{figure}[h]
\centering
  \includegraphics[width=9cm]{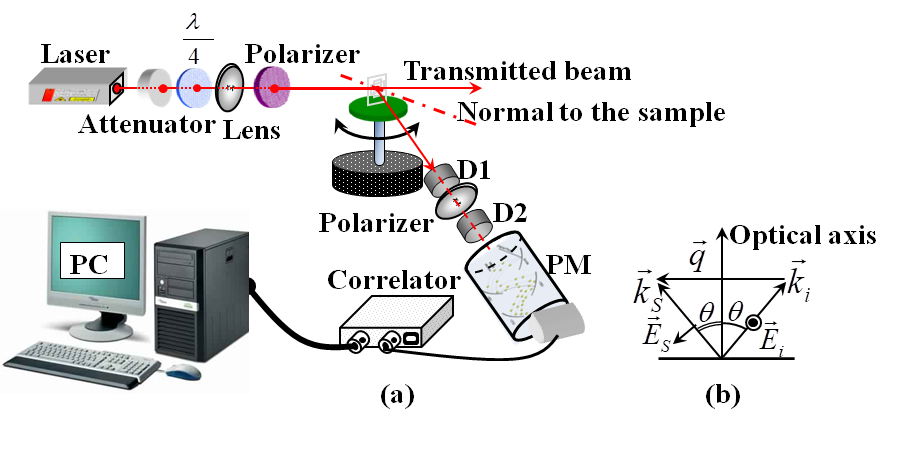}
  \caption{(a) Experimental set-up of the dynamic light scattering (DLS), 
(b) The scattering geometry, where the wave vector $k_{s}$ of the scattered light is quasi-symmetric to the wave vector $k_{i}$ of the incident light with respect to the optical axis (exactly symmetric in air), with an ordinary polarization for the incident light and an extraordinary polarization for the scattered light.
}
  \label{fgr:setup}
\end{figure}

The study of the viscoelastic behavior of nematic liquid crystal cells doped with nanoparticles was performed using dynamic light scattering (DLS). The light beating technique consists in measuring the autocorrelation function $C(t)$ of the photocurrent associated with light scattered by thermal fluctuations of the nematic director.
The cell was placed onto a goniometer stage that allows for scanning the incident angle in the range 3$^{\circ}$ to 15$^{\circ}$.  A vertically polarized laser beam passes through a convergent lens and then reaches the cell at a defined incidence angle $\theta$ (Figure \ref{fgr:setup}b). This allows us to increase the light intensity in the scattering volume and to reduce the number of the coherence areas on the sensitive surface of the photocathode of the detector.  \textcolor{blue}{We define the coherence number, $N$ as the ratio between the surface of the photocathode of the photomultiplier and a coherence area, which is the set of points for which the electromagnetic field of the light wave has the same phase (zero phase shift). In our case, this number is smaller than one (N $\approx$ 0.02)}.  The scattered light is collected through two diaphragms D1 and D2, both placed before of the detector (Figure \ref{fgr:setup}a).  The first diaphragm D1 selects the direction of the scattered wave vector and reduces the noise. The diaphragm D2 defines the number of coherence areas on the sensitive surface of the photo-detector. The diaphragms D1, D2 and the detector are placed on a rotating arm whose axis of rotation coincides with the goniometer axis (Figure \ref{fgr:setup}a).  The detection of the scattered light is carried out using a photomultiplier tube followed by a correlator that gives the auto correlation function which is then fitted with a theoretical model \cite{Othman95}. We work in a scattering geometry associated with a nearly pure twist mode. The wave vector $k_{s}$ of  the scattered light is symmetric to the wave vector $k_{i}$ of the incident light with respect to the optical axis in the air (equal incident and scattering angles in air $\theta$). This means that in air, the scattering vector $q$ is perpendicular to the optical axis. In the liquid crystal, $k_{i}$ and $k_{s}$ are not exactly symmetric but it can be shown that with an ordinary polarization for the incident light and an extraordinary polarization for the scattered light, the parallel component of the scattering vector $q$ can be neglected with respect to the perpendicular component, $q_{\perp}$ ($q_{||}/q_{\perp} < 2\times10^{-5}$) (Figure \ref{fgr:setup}b) \cite{Othman95}.

\begin{figure}[h]
\centering
  \includegraphics[width=9cm]{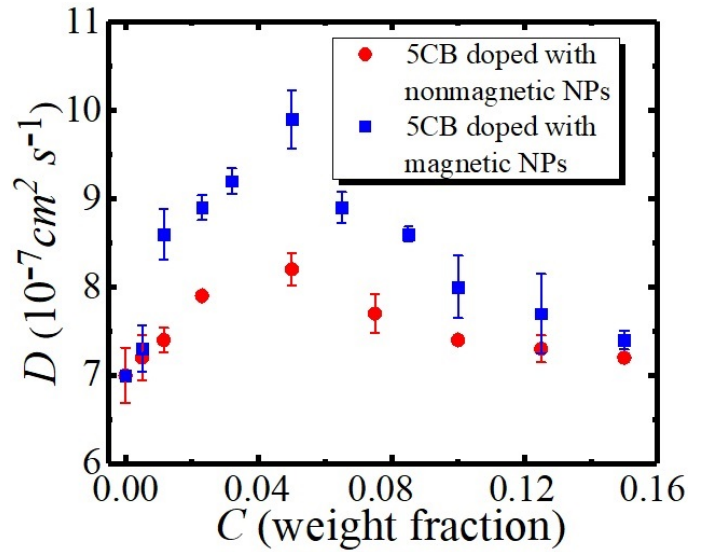}
  \caption{Evolution of the diffusion coefficient as a function of the nanoparticle weight concentration for non-magnetic and magnetic nanoparticles.}
  \label{fgr:dls}
\end{figure}

The fitting of the experimental autocorrelation function allows for the determination of the relaxation time of the twist deformation mode. For the fitting, the following equation has been used:
\begin{equation}
C(t) = A exp(-t/\tau) + B  exp(-2t/\tau) + C,
\end{equation}
where $\tau$ is the relaxation time corresponding to the collective excitations of the twist mode.
$\tau$ is experimentally shown to depend on the incident angle in air $\theta$, as it relates to the diffusion coefficient. $D = 1/(\tau q_{\perp}^{2})$ :
In the twist mode where $q_{||}$ can be neglected with respect to $q_{\perp}$, $\tau = \eta / (K_{2} q_{\perp}^{2})$, with $\eta$ the rotational viscosity, $K_{2}$ the twist elastic coefficient and $q_{\perp}$ the wave-vector transfer, perpendicular to the optical axis \cite{Groupe1969, Gennes1995,Othman95}. As a result, $D$ does not depend on the incident angle and can be extracted from the $\tau$ measurement using the $q_{\perp}$ value: $q_{\perp} = 4(\pi/\lambda) \sin(\theta)$. Figure \ref{fgr:dls} shows that $D$ initially increases when the concentration of nanoparticles increases, with a curve shape approximately inverse from the one of the magnetic Fredericksz threshold (Figure \ref{fgr:field_concentration}). After a maximum value, D starts to decrease when the concentration of both nanoparticles is increased above a critical concentration of the order of 0.05 weight fraction, i.e. analogous to the value at the minimum threshold field of the Fredericksz transition.

\section{Discussion}
\begin{itemize}
\begin{figure}[h]
\centering
  \includegraphics[width=8cm]{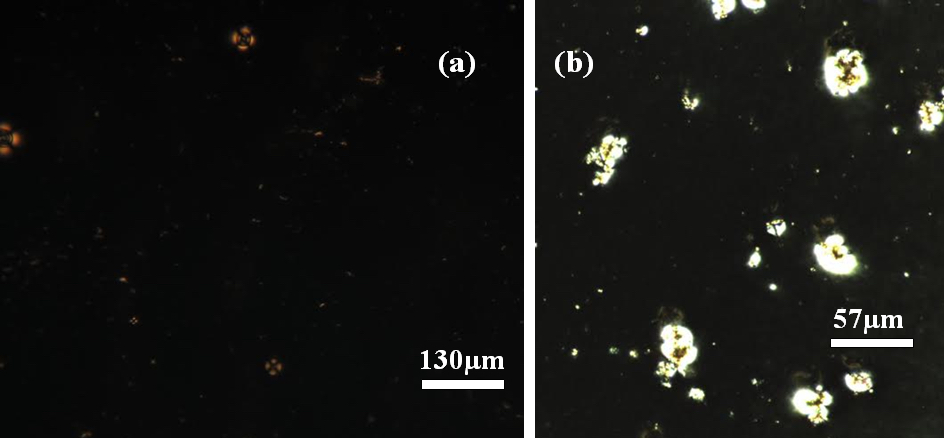}
  \caption{Optical Microscopy picture between crossed polarizer of a composite film 5CB/$Fe_{2}O_{3}$ with (a) 0.0115 weight fraction and (b) 0.023 weight fraction.}
  \label{fgr:MO}
\end{figure}
\item When the concentration $c$ increases, we observe a deacrease of $H_{c}$ down to a minimum at \textcolor{blue}{0.05} weight fraction, followed by a reversal of this trend for concentrations larger than 0.05 weight fraction. This observation can be interpreted by the presence of aggregates of nanoparticles.
Evidence of aggregation is indeed revealed by optical microscopy between crossed polarizers, already for concentrations equal to 0.02 weight fraction (Figure \ref{fgr:MO}). This observation is supported by the DLS measurements, demonstrating that the diffusion coefficient $D$ presents a maximum at the same concentration ($c = 0.05$ weight fraction) at which $H_{c}$ presents a minimum (Figure \ref{fgr:dls}). The result of aggregation is a slowing down of the appearance of new nanoparticle surfaces when the nanoparticle concentration increases. For $N$ nanoparticles assembled in a spherical aggregate, the surface area of the
aggregate scales as $N^{2/3}$ instead of being proportional to $N$ for isolated nanoparticles dispersed in the liquid crystal. The observation of an \textit{increase} of $H_{c}$ together with a \textit{decrease} of $D$ for concentrations larger than 0.05 weight fraction could be related to an acceleration of the aggregation process when the concentration increases above 0.05 weight fraction. Above this critical concentration of aggregates, a further increase of the nanoparticle concentration would increase the number of newly aggregated nanoparticles by more than the number of new individual nanoparticles, thus consuming isolated nanoparticles in the liquid crystal and finally leading to a quasi-disappearance of isolated nanoparticles. When the concentration of isolated nanoparticles decreases in favor of the aggregated nanoparticles, the total surface of nanoparticles exposed to the liquid crystal decreases and we observe a lower influence of the nanoparticles dispersed in the nematic matrix on the reorientation of the director along the magnetic field. Therefore a major role may be played by the total nanoparticle surface area on the observed phenomena, evolution of the magnetic Fredericksz threshold and the diffusion coefficient. The critical value of concentration $c = 0.05$ weight fraction can thus be viewed as a value below which the influence of single nanoparticle dominates and above which the influence of aggregate dominates. $c = 0.05$ corresponds to a volume fraction around $10^{-2}$, a particularly large value with respect to previous works describing an aggregation influence for smaller volume fractions \cite{Podoliak2012}. This may be due to the PDMS coating around the nanoparticles and the expected planar anchoring induced for 5CB (see below) that would lead to aggregation phenomena for large concentrations only, in agreement with recent data where nanoparticles have been grafted with mesogene molecules of dendritic structure \cite{Prodanov2016}.

\begin{figure}[h]
\centering
  \includegraphics[width=8cm]{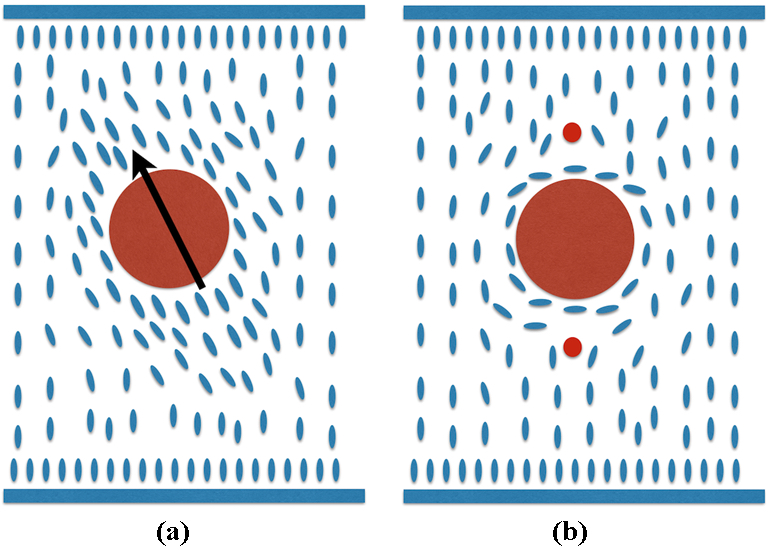}
  \caption{Scheme of the expected liquid crystal geometry around (a) iron oxide ($\gamma Fe_{2}O_{3}$) nanoparticles with a magnetic moment inducing a planar anchoring with a director orientation parallel to the magnetic moment of the nanoparticle in its vicinity. This is possible due to the spherical shape of the nanoparticle, here represented in side view; (b) cerium oxide ($CeO_{2}$) nanoparticles without magnetic moment and with planar anchoring of the director. We expect formation of defects roughly indicated by the red points.}
  \label{fgr:fig8}
\end{figure}
\item We now focus on the small concentration regime where single nanoparticles dominate. The first result is that cerium oxide nanoparticles, functionalized with PDMS, modify the magnetic Fredericksz transition with a decrease of the Fredericksz threshold, $H_{c}$,  despite their non-magnetic character, whereas, to the best of our knowledge, up to now only low magnetic field features had been shown for non-magnetic nanoparticles \cite{Tomasovicova2013}. $H_{c}$ decreases continuously when the concentration of nanoparticles increases, until the limiting concentration value, $c = 0.05$ weight fraction, at which point $H_{c}$ has decreased by approximately 23 \%.
For a nematic liquid crystal film with rigid anchoring at the substrates, the magnetic field threshold is $H_{c} = (\pi/e)(K_3/\chi_a)^{1/2}$, where $K_3$ is the 5CB bend elastic constant and $\chi_a$ is the 5CB magnetic susceptibility anisotropy. $K_3$ is known to be proportional to $S^2$, $S$ being the scalar orientational order parameter \cite{Rosenblatt1984}. Since $\chi_a$ is proportional to $S$  \cite{Gennes1995}, $H_{c}$ is proportional to $S^{1/2}$. From the initial decrease of $H_{c}$ with concentration, we might infer that there is a decrease in the spatially averaged order parameter $S$ due to the presence of nanoparticles. However DLS measurements appear to contradict this assumption. The diffusion coefficient $D = K_2 / \eta$ increases when the concentration of nanoparticles increases, with a curve shape inverse with respect to that of the magnetic Fredericksz threshold, reaching a maximum increase of 13\% for $c = 0.05$ weight fraction. Similar to $K_3$, $K_2$ is proportional to $S^2$  \cite{Karat1977}. $\eta$ is proportional to $a S + b S^{2}$, where $a$ and $b$ are constants and the overall proportionality involves an activation energy \cite{Prost1976, Belyaev1985, CHMIELEWSKI1986, Blinov1994, Wu1990, Belyaev2001}. As a result, we expect $D$ to increase with $S$ , suggesting an initial \textit{increase} of the order parameter in the hybrid film with nanoparticle, when the nanoparticle concentration increases. These two apparently contradictory behaviors lead to the conclusion that the average order parameter may not vary significantly with $c$; instead the coefficients defining the proportionality of the different physical parameters with the order parameter may vary with  $c$. $K_3 = f(c) S^{2}$, the assumption being that bend and twist elastic constants behave similarly, $\chi_a = g(c) S$ and $\eta = h(c) (a S + b S^{2})$. \textcolor{blue}{$a$ and $b$ defining the respective weights of the $S$ and $S^{2}$ terms, it is not obvious that thay would significantly vary with the concentration}. With non-magnetic nanoparticles, we do not expect that $g(c)$, related to $\chi_a$, significantly varies. The decrease of $H_{c} = (f(c) S/g(c))^{1/2}$ thus indicates that the product $f(c) S$ decreases when $c$ increases. This shows that the elastic constant itself decreases when the concentration of nanoparticle increases, which may be related to a planar anchoring on the nano particles covered by PDMS. Figure \ref{fgr:fig8} shows the two schematized geometrical pictures of the liquid crystal molecules in the vicinity of the  magnetic (Figure \ref{fgr:fig8}a) and non-magnetic (Figure \ref{fgr:fig8}b) nanoparticles. For non-magnetic nanoparticles planar anchoring may lead \textcolor{blue}{on the one hand} to defects at the nanoparticle poles. However it is known that the defects are clearly defined for particles of micrometer size. For nanometer sizes, the defect nature is less clear and we only underline in red their expected location on Figure \ref{fgr:fig8}b \cite{Tomar2012, blanc2013}.  A planar anchoring on the PDMS brush may facilitate \textcolor{blue}{on the other hand} the nematic director bend \textcolor{blue}{for nanoparticles of spherical shape}, if the anchoring is weak enough. \textcolor{blue}{From} the point of view of anchoring, a tilted molecular orientation in the vicinity of the nanoparticles under the influence of the external magnetic field indeed remains equally favorable \textcolor{blue}{compare to} the initial state without external magnetic field. \textcolor{blue}{This may lead to a decrease of the average bend elastic constant}.

However the diffusion coefficient $D = K / \eta$, being equal to$f(c) S^{2}/h(c) (a S + b S^{2})$, has been shown to increase when $c$ increases, despite the decrease of $f(c) S$. This leads to the conclusion that if the order parameter, $S$, remains approximately constant, $h(c)$, related to the viscosity, may decrease even more rapidly when $c$ increases. Here, we can consider that $(f(c) S(c))/(f(0) S(0)) = (H_{c}/H_{0})^2 = 0.59$, for $c = 0.05$. This could also correspond to a decrease of the bend elastic constant by 0.59 if $S$ does not vary. Concerning viscosity, we obtain $\eta (c)/\eta(0)$ equal to $h(c) (a S + b S^{2})(c)/h(0) (a S + b S^{2})(0) = (D(0) f(c) S^{2}(c))/(D(S) f(0) S^{2}(0)) = (0.59/1.13) (S(c)/S(0))$. This is equal to 0.52 if the order parameter, $S$, can be considered as constant. This corresponds to a relative decrease of viscosity of about one half. It is indeed known that unlike for simple fluids, the viscosity of liquid crystals decreases in the presence of nanoparticles \cite{Miyamoto2009, vardanyan2012}. The phenomenon obviously depends on nanoparticle nature, concentration, size and ligands. \textcolor{blue}{In presence of} metallic nanoparticles covered by 5CB, a closely related decrease of the \textcolor{blue}{5CB} viscosity by one third, \textcolor{blue}{close to the one found above} has been found \cite{Miyamoto2009}, but for gold nanoparticles of diameter  10nm, a decrease by about 4.6 times has been found \cite{vardanyan2012}. We finally can conclude that, qualitatively, for non-magnetic nanoparticles, the $H_{c}$ evolution is related to the planar anchoring induced by PDMS together with the spherical shape of the nanoparticles, which may favor an easier rotation of the liquid crystal director in their vicinity in the presence of an external magnetic field. It is interesting to notice that similar behaviors may occur with electrical Fredericksz transitions as well, together with the already discussed charge effects \cite{Urbanski2016}. The presence of nanoparticles also may favour a significant decrease of the the 5CB viscosity, which has been shown here by DLS experiments.
\begin{figure}[h]
\centering
  \includegraphics[width=8cm]{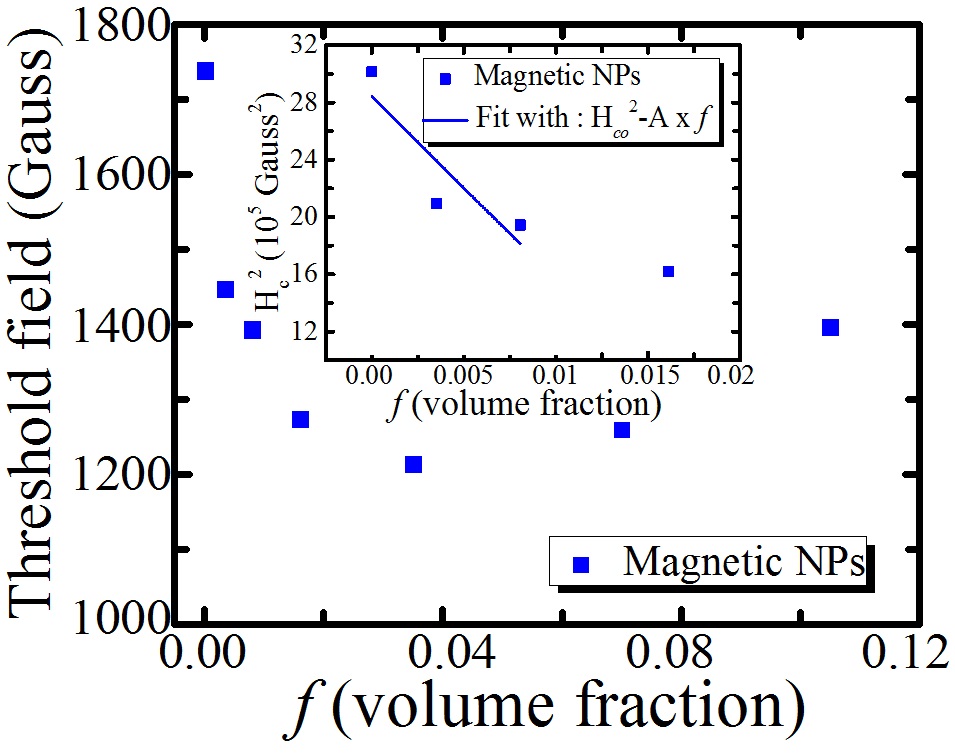}
  \caption{Evolution of the magnetic threshold field as a function of the nanoparticle volume fraction, $f$, for magnetic nanoparticles. The dashed lines represents the best fit of the function $H_{c}^{2}= H_{co}^{2} - A \times f$ in the range of volume fraction $f$ belonging to [0; 0.01] . }
  \label{fgr:fit}
\end{figure}
\item For magnetic nanoparticles, $H_{c}$ values below those of non-magnetic nanoparticles are observed (Figure \ref{fgr:field_concentration}), but the difference between the two kinds of particles remains small (31\% versus 23\% \textcolor{blue}{decrease of $H_{c}$}, whereas a much larger difference in the evolution of the diffusion coefficient $D$ is observed (Figure \ref{fgr:dls}). It is remarkable that this occurs for similar sizes and similar PDMS coatings of both particles, in principle providing the same liquid crystal geometry in the vicinity of both kind of nanoparticle. This suggests that the magnetic character plays a significant role in the induced $D$ value, which can be understood based on the scheme of Figure \ref{fgr:fig8}a. It shows how the geometry of liquid crystal molecules around the nanoparticles can be significantly modified if the magnetic interactions between nanoparticles and liquid crystal molecules are non-negligible. Indeed, for the PDMS shell in a poor solvent, being of thickness of order of 2nm, we expect at the nanoparticle surface a magnetic field of roughly 6kG (dipolar field approximation at 2 nm from the surface of a nanoparticle with a magnetic moment $\mu=m_sV_p$, $m_s$ being the saturation magnetization of the ferrite and $V_p$ the volume of the nanoparticle), which might result in a  fixed orientation of the liquid crystal molecules parallel to the magnetic dipole in the vicinity.  The corresponding anchoring on the PDMS-covered nanoparticles may remain planar since all orientations parallel to the magnetic moment are consistent with a planar anchoring due to the spherical shape of the nanoparticles. This is true everywhere around the nanoparticles except close to the nanoparticle poles where we expect an homeotropic anchoring due to the dominant influence of the nanoparticle dipolar magnetic field (Figure \ref{fgr:fig8}a). The length along which liquid crystal molecules are oriented parallel to the nanoparticle magnetic dipole is not easy to estimate due to the small size of the nanoparticles with respect to the nematic extrapolation length. Moreover, due to their superparamagnetic behavior, each nanoparticle, in the absence of external magnetic field, may present a fluctuating and disordered magnetic dipole moment. It is thus not clear if this would lead to an also fluctuating  region of liquid crystal molecules around the nanoparticle (Figure \ref{fgr:fig8}a) or if the coupling between nanoparticle dipole moment and the nematic director would be large enough to orient and stabilize the magnetic dipole of the nanoparticles parallel to the average director orientation \cite{Brochard1970_JP, Mertelj2013}. Due to the difference by 3 orders of magnitude between the dipole fluctuation time (the so-called N\'eel time in the ns-$\mu$s range) and the 5CB director modes detected by DLS (in the ms range), we do not expect a significant impact of the N\'eel fluctuations on the measured 5CB coefficient $D$, with respect to non-magnetic nanoparticles. The large $D$ increase may reflect mainly an increase of the average order parameter in the composites \textit{with} magnetic nanoparticles, as compared to the composites with non-magnetic nanoparticles This order parameter increase should in principle also lead to an increasing $H_{c}$. However this increase of $H_{c}$ is not observed showing that the magnetic response of the system with magnetic nanoparticles is of different nature with respect to the one with non-magnetic nanoparticles. This is clearly due to the expected reorientation of the iron oxide nanoparticle dipoles in the presence of external magnetic field, which in turn, may rotate the surrounding "shell" of liquid crystal molecules well oriented around each nanoparticle. The fact that the $H_{c}$ decrease occurs for nanospheres in relation with planar anchoring also is consistent with the recent observation of a similar decrease of $H_{c}$ in E7 liquid crystal with cobalt ferrite nanoparticles with ligands made of mesogene molecules of dendritic structure, also associated with planar anchoring \cite{Prodanov2016}.   \textcolor{blue}{A decrease of $H_{c}$ was predicted for planar anchoring of nanorods \cite{Zakhlevnykh2014, Zakhlevnykh2016}. However simple theory allowing for a direct comparison with experimental data now available, in particular as a function of the nanorod concentration, would be useful. We notice that the observed decrease of the magnetic Fredericksz threshold can be fitted by a curve $H_{c}^{2}= H_{co}^{2}- A  \times f$ (Figure \ref {fgr:fit}) for the first points of the curve, $f$  being the volume fraction instead of the concentration and $H_{co}$ the magnetic Fredericz threshold of pure 5CB. This fitting is similar to the one used in ref. \cite{Kopcansky2013} whereas in this latter case nanorods instead of nanospheres were used. It only differs by the sign of the evolution with respect to the calculated increase of magnetic threshold for magnetic nanorods with homeotropic anchoring at low concentrations ($H_{c}^{2}= H_{co}^{2} + 2Wf/(\chi_a d)$ with W the anchoring energy at the nematic-magnetic particle boundary and d the size of the particles \cite{Burylov1993}.
By fitting our results using $H_{c}^{2}= H_{co}^{2}- A \times f$ we found $A  \approx 1.5 \times 10^{4} (SI)$. P. Kop{\v{c}}ansk{\`y} et al. found for low concentrations of nanorods made of a mixture of magnetite and hematite a value $A \approx 5.6 \times 10^{3} (SI)$ for nanorods of length 50nm, diameter 10nm and $A \approx 12 \times 10^{3}(SI)$ for longer nanorods of length 400nm, diameter 18nm.
When an increase is predicted for homeotropic anchoring, the slope of the increase with volume fraction is $A = 2 \mu_o W/(\chi_a d)$, leading to an order of magnitude for  $A \approx  3.6 \times 10^{2} (SI)$ \cite{Burylov1993}. The slopes of decrease for nanospheres and nanorods are thus of the same order and also of the same order as the calculated slope of the increase for homeotropic anchoring on nanorods. However we expect a decrease of the slope when the nanoparticle diameter $d$ is increased but the contrary behavior is found. This could suggest that for planar anchoring a different evolution of $H_{c}$ with respect to the concentration would occur compared to the case of homeotropic anchoring.}

  \textcolor{blue}{Our results finally suggest that the behavior of non-magnetic and magnetic nanoparticle composites with 5CB nematic matrix can be intrinsically different, with the decrease of $H_{c}$  being more efficient with the use of magnetic nanoparticles.} For both kinds of nanoparticle, the planar anchoring driven by PDMS appears crucial. In a previous work with magnetic nanoparticles of analogous shape and size but coated by a a  mixture  of  mono-  and  diesters  of  phosphoric acid, the dispersion in 5CB was possible only in the isotropic phase (up to 1 \% volume fraction), phase separation occurring in the nematic phase \cite{Cruz2005}. Here we obtain a true ferronematics with individually dispersed non-magnetic or magnetic nanoparticles in the nematic 5CB matrix up to 1\% volume fraction and a significant decrease of $H_{c}$ similar to the $H_{c}$ decrease obtained for coating cobalt ferrite nanoparticles with ligands made of mesogene molecules of dendritic structure, also associated with planar anchoring in E7 liquid crystal \cite{Prodanov2016}. \textcolor{blue}{Our results thus confirm the key role played by the liogands to obtain, not only a decrease of $H_{c}$ (planar anchoring necessary) but also a large volume fraction of nanoparticles without aggregation}.However, we show that for magnetic nanoparticles, the coupling between the orientation of the magnetic moment and orientation of the nematic director is driven not only by the weak planar anchoring on PDMS ligands, but also by magnetic interactions between liquid crystal molecules and nanoparticles; these are usually neglected in the theoretical models.  The curve of the magnetic Fredericksz threshold for magnetic nanoparticles with planar anchoring as a function of concentration, allowing for an initial  decrease of the magnetic Fredericksz threshold, has rarely been obtained up to now, except in the liquid crystal MBBA \cite{Kopcansky1997_JP}. However in this last case the two zones of the curve, one with limited aggregation (low concentration), the other where aggregation dominates the value of $H_{c}$, were not clear. Here the interpretation of the first zone of the $H_{c}$ and D curves (respectively decreasing and increasing with concentration) were qualitatively ascribed to the influence of single nanoparticles dominating over aggregates, with a clear  decrease of magnetic Fredericksz threshold. \textcolor{blue}{This would now require a theory to extract the relevant parameters characterizing the nanoparticles and their magnetic influence in the specific situation where nanospheres with planar anchoring are involved}.
 
\end{itemize}

\section{Conclusion}
In conclusion, we have found that a PDMS coating efficiently leads to planar anchoring on magnetic and non-magnetic nanospheres within a 5CB liquid crystal matrix, allowing for a net decrease of the magnetic threshold (from 23\% to 31\%) for nanoparticle concentration small enough at room temperature. We show that non-magnetic nanoparticles significantly reduce the magnetic Fredericksz threshold, in line with a decrease of the average 5CB elastic constants. An even larger decrease of the viscosity, by around a half of the initial value, is shown by DLS data. We show that \textit{magnetic} nanoparticles present an only slightly larger decrease of the threshold compared to non-magnetic nanoparticles (31\% versus 23\%), but the phenomenon at the origin of this decrease appears to be quite different from the case of non-magnetic nanoparticles.  A significant reorientation of the liquid crystal molecules occurs around the magnetic nanoparticles driven by the nanoparticle magnetic dipole, in line with an increase of the order parameter that may become larger, on average, than for non-magnetic nanoparticles. The resulting decrease of the elastic constants may thus be smaller, but the non-negligible local magnetic interactions between liquid crystal molecules and magnetic nanoparticles favor an easier liquid crystal rotation in the presence of external magnetic field. The applied magnetic field is able to reorient the magnetic moments, along with the mesogene molecules. For both kinds of nanoparticles, the curve of the magnetic Fredericksz threshold when nanoparticle concentration increases has been obtained, allowing for an accurate analysis of two distinct  zones: the first zone at low concentration (concentration lower than 0.05 weight fraction  (or volume fraction lower than $ 3 \times 10^{-2}$), where the influence of individually dispersed nanoparticles may dominate, the second zone at high concentration, where aggregation is more rapid than the production of isolated nanoparticles. This curve confirms that isolated nanoparticles are more efficient in decreasing the magnetic threshold. The decrease of the threshold can be fitted by \textcolor{blue}{a curve $H_{c}^{2}= H_{co}^{2}- A \times f$, but the evolution of $A$ with the nanoparticle diameter is inverse from the one predicted for homeotropic anchoring and $H_{c}$ increase}. Our results now need a theoretical interpretation, involving both planar anchoring and spherical nanoparticles, in line with the observed reorientation of the liquid crystal molecules driven by the magnetic dipole moment of the magnetic nanoparticles.

 %For footnotes in the main text of the article please number the footnotes to avoid duplicate symbols. e.g.  \footnote[num]{your text} the corresponding author \ast counts as footnote 1, ESI as footnote 2, e.g. if there is no ESI, please start at [num]=[2], if ESI is cited in the title please start at [num]=[3] etc. Please also cite the ESI within the main body of the text using \dag.

\section{Acknowledgement}
IRN, JSP, and CR were funded by the National Science Foundation's Condensed Matter Physics program under grants DMR-1065491 and DMR1505389.Travel between Paris and Cleveland was supported by the Partner University Fund and the work was also partly funded by UPMC-Matisse labex. We acknowledge Brigita Rozic for useful advices for the preparation of the composites.


%\balance


\bibliography{biblio.bib}

\end{document}